\documentclass{article}

\usepackage{arxiv}

\usepackage[utf8]{inputenc} 
\usepackage[T1]{fontenc}    
\usepackage{hyperref}       
\usepackage{url}            
\usepackage{booktabs}       
\usepackage{amsfonts}       
\usepackage{nicefrac}       
\usepackage{microtype}      
\usepackage{lipsum}     
\usepackage{graphicx}
\usepackage[style=apa, natbib=true, backend=biber]{biblatex}
\usepackage{doi}

\newcommand{\Autoref}[1]{%
  \begingroup%
  \def\chapterautorefname{Chapter}%
  \def\sectionautorefname{Section}%
  \def\subsectionautorefname{Subsection}%
  \autoref{#1}%
  \endgroup%
}

\title{Incorporating Blogs in Pollux}

\author{Tobias Holtdirk\thanks{Corresponding author}\\
    GESIS -- Leibniz Institute for the Social Sciences\\
    Unter Sachsenhausen 6-8 \\
    50667 Köln \\
    \texttt{tobias.holtdirk@rwth-aachen.de}
    \And 
    Nina Smirnova\\
    GESIS -- Leibniz Institute for the Social Sciences\\
    Unter Sachsenhausen 6-8 \\
    50667 Köln \\
    \texttt{nina.smirnova@gesis.org}
}

\renewcommand{\headeright}{Technical Report}
\renewcommand{\undertitle}{Technical Report} 
\renewcommand{\shorttitle}{Incorporating Blogs in Pollux}

\hypersetup{
pdftitle={Technical Report: Incorporating Blogs in Pollux},
pdfsubject={cs.DL},
pdfauthor={Tobias Holtdirk, Nina Smirnova},
pdfkeywords={political blogs, data crawling, political science, social media},
}

\begin{document}
\maketitle

\begin{abstract}
    This technical report describes the incorporation of political blogs into Pollux, the Specialised Information Service (FID) for Political Science in Germany. Considering the widespread use of political blogs in political science research, we decided to include them in the Pollux search system to enhance the available information infrastructure. We describe the crawling and analyzing of the blogs and the pipeline that integrates them into the Pollux system. To demonstrate the content of the incorporated blogs, we also provide a visualization of the topics covered by the blog posts during the first three months following integration.
\end{abstract}


\section{Introduction}
Political blogs are widely used in political science research \citep{wallsten_agenda_2007, coleman_political_2008, wallsten_political_2008, guner_political_2009, akinnubi_deliberative_2023, peng_role_2023}. \citet{coleman_political_2008} claim that communication through political blogs might enhance the responsibility and openness of governance overall. \citet{peng_role_2023} examined the process of opinion formation and evolution by analyzing a larger social network of political blogs. \citet{wallsten_political_2008} investigated how political bloggers utilize their platforms, focusing on their predominant function: expressing opinions, mobilizing, seeking feedback, or disseminating information. \citet{demasi_analysing_2020} studied the usage of political blogs by right-wing politicians for political communication and persuasion. \citet{balakhonskaya_communicative_2020} examined strategies for discrediting opponents in the Russian political blogosphere.

Considering the broad usage of political blogs in political science research, we decided to include political blogs in the search index of Pollux \footnote{\url{https://www.pollux-fid.de/}}. Pollux is the Specialised Information Service (FID) for Political Science, which provides literature and information infrastructure in the field of political science in Germany. The following technical report explains the crawling procedure and analysis of political blogs we included in our collection. \Autoref{sec:sourcing} explains the generation of the list of RSS feeds used for incorporating blogs in Pollux. \Autoref{sec:analysis} describes findings from the collected RSS feeds that motivated the decision of how to implement them into Pollux. \Autoref{sec:incorporation} provides a detailed description of the procedure that handles the integration of blogs in Pollux. \Autoref{sec:Visualization} shows visualizations of the topics included in the incorporated blog posts.

\section{Sourcing of RSS Feeds}\label{sec:sourcing}
\begin{figure}[htb]
    \centering
    \includegraphics[width=.7\textwidth]{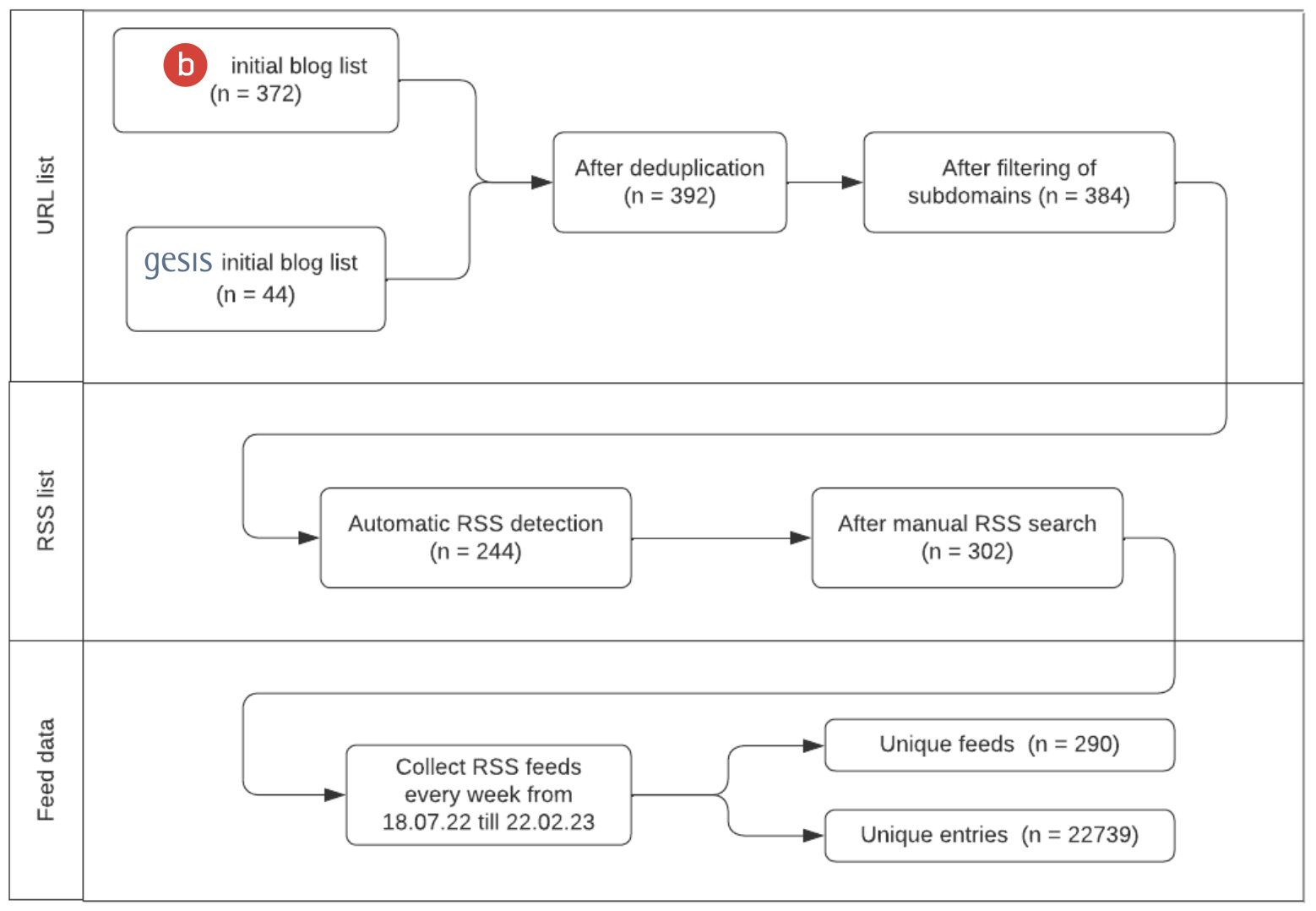}
    \caption{Pipeline for initial blog list generation.}
    \label{fig:rss_generation}
\end{figure}

In this section, we describe the generation of the RSS feeds list used to incorporate blogs in Pollux. A graphical overview of the process is provided in \autoref{fig:rss_generation}.

We use lists of URLs with relevant political science blogs provided by SUUB and GESIS. The lists are parsed, tested for dead links, and then filtered to deduplicate and fix subdomain inconsistencies between the lists.

To get an RSS list from the URL list, we designed a Python script for identifying RSS feeds on web pages. It reads a list of URLs from a CSV file, sends a GET request to each URL, and uses a regular expression to find any RSS feed links in the HTML of the page. The regular expression is designed to match the \texttt{href} attribute of link elements with \texttt{rel="alternate"} and \texttt{type="application/rss+xml"}, which is a common way to specify an RSS feed on a webpage. Finally, a dictionary containing the original URL, the response status code, the content type of the response, and the list of RSS feed links is written to a JSON file, which is used for the analysis of the RSS feeds, as well as the later integration into the Pollux database.

\begin{figure}[htb]
    \includegraphics[width=.7\textwidth]{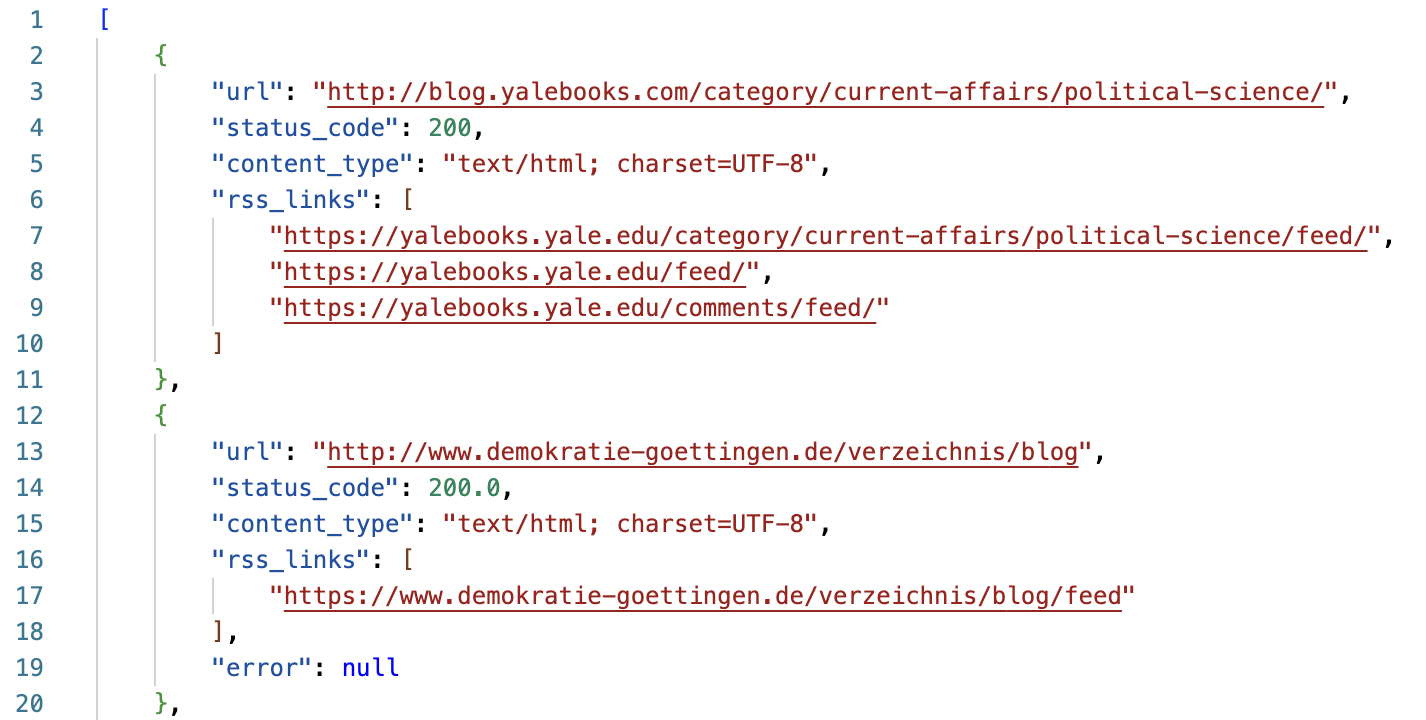}
    \caption{Two entries in JSON after retrieving RSS feeds from URLs.}
    \label{code:rss_list_json}
\end{figure}

After the automatic RSS feed detection, the remaining feeds are checked manually. Like in the automatic checking, the HTML of the URL is used. We look for broader patterns like an "rss" string appearing in a link on the page. After that, we check the remaining URLs by visiting the website and searching for visual information like an RSS feed logo. The additional RSS feeds identified this way are added to the automatically detected ones, providing a JSON file as seen in \autoref{code:rss_list_json}.

To analyze the feeds later, we downloaded the RSS feeds each week for seven months, from July 2022 to February 2023. We designed a Python script for this purpose that was executed once per week. The script downloads and saves RSS feeds from the list generated earlier. The feed content and metadata, such as the timestamp, status code, and content type, were saved. If an error occurs during a request, the request is retried multiple times.

\section{Analysis of Initial Blog Data}\label{sec:analysis}

After generating the initial feed data, we analyzed the resulting 290 feeds and 22,739 entries to identify which metadata is available and assess the quality of the available metadata.

For an overview of the feed-level data we parsed from the RSS feed, see \autoref{tab:feed_structure}.
At the feed level, only a little metadata is transmitted. However, the available metadata is included in almost all feeds and can, therefore, be used as information in Pollux. The fields mostly contained high-quality data, with only the \textit{subtitle} field showing some inconsistencies, as some feeds had considerably longer entries than others. Therefore, we were able to use all the shown field data in the Pollux integration.
\begin{table}[htb]
    \caption{\textbf{Feed Structure.} XML elements (fields) that were most often included in the RSS responses at the feed level.}
    \centering
    \begin{tabular}{lrl}
        \toprule
        \cmidrule(r){1-2}
        \textbf{Field} & \textbf{Inclusion} & \textbf{Example}                                   \\
        \midrule
        title          & 100\%              & \verb|PoliSciZurich                              | \\
        subtitle       & 99\%               & \verb|A blog by political scientists in Zurich   | \\
        blog url       & 100\%              & \verb|https://poliscizurich.wordpress.com        | \\
        rss url        & 100\%              & \verb|https://poliscizurich.wordpress.com/feed/  | \\
        last updated   & 92\%               & \verb|Wed, 17 Aug 2016 03:54:00 +0000            | \\
        language       & 89\%               & \verb|en-US|                                       \\
        \bottomrule
    \end{tabular}
    \label{tab:feed_structure}
\end{table}

For an overview of the entry-level data we parsed from the RSS feed, see \autoref{tab:entry_structure}. The entry level has more metadata available, sometimes even including the whole blog post in the content field. However, the quality and availability of data are worse. The \textit{title} is always available and has mostly good quality, with some titles being overly long and some not including the actual title but rather a "not available" statement. Important metadata like \textit{link} and \textit{publication date} also had good quality. Other metadata was more spotty; \textit{content} often includes HTML artifacts and had inconsistent content, with some entries having their entire blog post in the field and others just the first sentence. On the other hand, the \textit{summary} field is available for all entries and has more consistent quality. Therefore, we use the summary rather than the content as the displayed "abstract" in Pollux. The \textit{tags} and \textit{comments} fields provide interesting metadata. With tags, the blog entries can be categorized into different subsets, and comments can be used to build some kind of popularity metric. However, comments have a very low inclusion rate at 33\% and are therefore not reliable enough for a metric. The amount and kind of tags differ heavily for different entries, making them less valuable than an automatic topic generation based on the summary.
\begin{table}[htb]
    \caption{\textbf{Entry Structure.} XML elements (fields) that were most often included in the RSS responses at the entry level.}
    \centering
    \begin{tabular}{lrl}
        \toprule
        \cmidrule(r){1-2}
        \textbf{Field}   & \textbf{Inclusion} & \textbf{Example}                                          \\
        \midrule
        title            & 100\%              & \verb|The 2022 Midterms: In the Senate elections, [...] | \\
        id               & 100\%              & \verb|https://blogs.lse.ac.uk/usappblog/?p=47109        | \\
        link             & 100\%              & \verb|https://blogs.lse.ac.uk/usappblog/2022/11/  [...] | \\
        publication date & 96\%               & \verb|2022-11-16 09:57:58|                                \\
        authors          & 84\%               & \verb|Blog|                                               \\
        summary          & 100\%              & \verb|In this year\&\#8217;s midterm elections,   [...] | \\
        content          & 65\%               & \verb|a class="a2a_button_twitter" href="https: [...] |   \\
        tags             & 65\%               & \verb|2022 Midterms', 'Elections and party politi [...] | \\
        comments         & 33\%               & \verb|https://blogs.lse.ac.uk/usappblog/2022/11/16[...] | \\
        \bottomrule
    \end{tabular}
    \label{tab:entry_structure}
\end{table}

Since we have high-quality data both at the feed and entry levels, we can use them for two distinct types of records in Pollux. As Pollux is designed as a database for academic research, we can use the preexisting structure of paper and journal records as a template for entry and feed records.

\section{Incorporation into the Pollux Pipeline}\label{sec:incorporation}

This section provides a detailed description of the pipeline that handles the integration of blogs in Pollux.

\subsection{Downloading RSS Feeds}

The Python script, \texttt{rss\_downloader.py}, is designed to download and store RSS feeds from a list of URLs. It does this through a series of functions that each handle a specific part of the process.

The \texttt{run} function is the main function that gets called to initiate the process. It reads a list of RSS feed URLs, shuffles the list to avoid querying the same domain consecutively, and then attempts to download each feed. If a feed is successfully downloaded, its metadata is stored. If a feed cannot be downloaded, its URL is added to a list of error URLs. The function then saves the metadata of all downloaded feeds to a JSON file and returns a \texttt{Result} object containing metrics about the number of feeds downloaded and the number of errors, as well as the error URLs.

Multiple helper functions are used in the process. The \texttt{extract\_feed\_urls} function reads in a JSON file containing blog information and extracts a list of unique RSS feed URLs. It expects the JSON file to contain a list of dictionaries, each with a key \texttt{rss\_links} that maps to a list of URLs. The function returns the list of unique URLs.

The \texttt{load\_rss} function handles the HTTP request that attempts to download an RSS feed from a given URL and save it to a specified directory. It first sends a GET request to the URL. If the response's content type is not XML, the function raises a ValueError. Otherwise, it saves the response's content to a file in the specified directory and returns a dictionary containing the URL, the timestamp of the download, the filename, the status code of the response, and the content type.

The \texttt{get} function handles the GET request used above. It sends a GET request to a given URL with specified headers and URL parameters. If the request is successful and the status code of the response is OK, the function returns the response. If the status code is a client error (400-499), the function raises a ValueError. If the status code is a server error (500-599), the function waits for a certain amount of time and then tries to send the request again. If the request fails three times, the function raises a \texttt{ValueError}.

\subsection{Converting RSS Feeds}

The Python script, \texttt{rss\_converter.py}, is designed to parse and convert RSS feeds and their entries into the format used by Pollux entries. It does this through a series of functions that each handle a specific part of the process.

The \texttt{run} function is the main function that gets called to initiate the process. It reads in a dump of RSS feeds, parses the feeds and their entries, splits the feeds and entries into blog and comment data, converts the blog feeds and entries into a specific format, and returns the converted data along with some metrics and logs.

The \texttt{parse\_rss\_dump} function parses the RSS feeds and their entries. It adds additional information from the metadata to each feed and entry.
The \texttt{split\_comments} function separates the blog data from the comment data based on the URL of the feed or entry.

The \texttt{convert\_feed} and \texttt{convert\_entry} functions convert a feed or an entry into the Pollux entry format. They call several helper functions to get specific pieces of information from the feed or entry.
Ten different helper functions in the format \texttt{get\_<field>} parse the blog information. For example, the \texttt{get\_languages} function converts the BCP 47 language code found in the RSS to an ISO 639-3 language code. For detailed information on the functions, see the accompanying source code.

\section{Topics of incorporated Blog Posts}\label{sec:Visualization}
After the integration of the pipeline into Pollux (see \autoref{appendix:website} and \citet{czolkoss-hettwerPolitikwissenschaftlicheBlogsSichtbar2023} for the website layout), we analyze the incorporated blogs. Integration started in July 2023, and we analyzed blog entries until October 2023. We analyze the topics of the blog entries by fitting a topic model on the \texttt{summary} field of the entries. Before fitting the model, we translated all summaries into English to make the topic model consistent, as we want to focus on the areas the blogs cover instead of language differences. We additionally use the \texttt{publication date} field to visualize topics over time.

\paragraph{BERTopic.}
BERTopic \citep{grootendorstBERTopicNeuralTopic2022} is a topic modeling technique that combines BERT (Bidirectional Encoder Representations from Transformers) with classical topic modeling methods. It leverages the power of BERT embeddings to represent documents and then applies topic modeling algorithms to discover latent topics within the document collection.
BERTopic provides several methods for visualizing the discovered topics and the documents associated with them. We visualize document-level relationships and topics over time.

\begin{figure}[!htb]
    \includegraphics[width=1.0\textwidth]{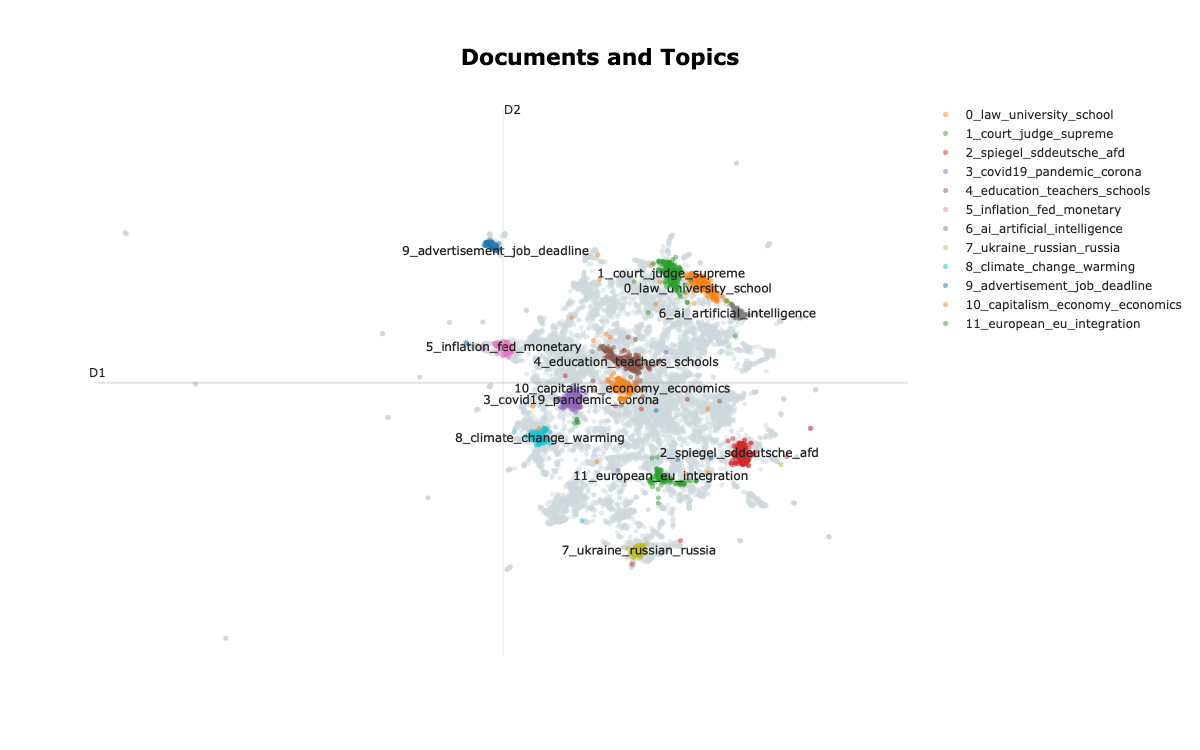}
    \caption{BERTopic visualization of blog entries incorporated in Pollux. The scatterplot shows embeddings for each blog entry reduced to a 2-dimensional space. The summaries of the entries were used to create the embeddings. (Interactive plot at \url{https://tobihol.github.io/pollux-rss-blogs/content_analysis/topic_viz_2d.html})}
    \label{fig:topic_viz_2d}
\end{figure}

\paragraph{Document visualization.}
Using the fitted topic model, we visualize the documents in the context of the discovered topics, see \autoref{fig:topic_viz_2d}. We plot the documents in a two-dimensional space, representing each as a point. The position of the documents is determined by their topic distribution, such that documents with similar topic distributions will be clustered together. This visualization helps understand the relationships between documents based on their topic assignments. We can see that US-centric topics like the Supreme Court, monetary policy, and AI advances dominate the visualization's upper half. The middle includes topics such as the coronavirus and climate change, which matter globally. The bottom half includes topics concerning the European Union, German newspapers, and the war in Ukraine, which are more relevant in Europe.

\begin{figure}[!htb]
    \includegraphics[width=1.0\textwidth]{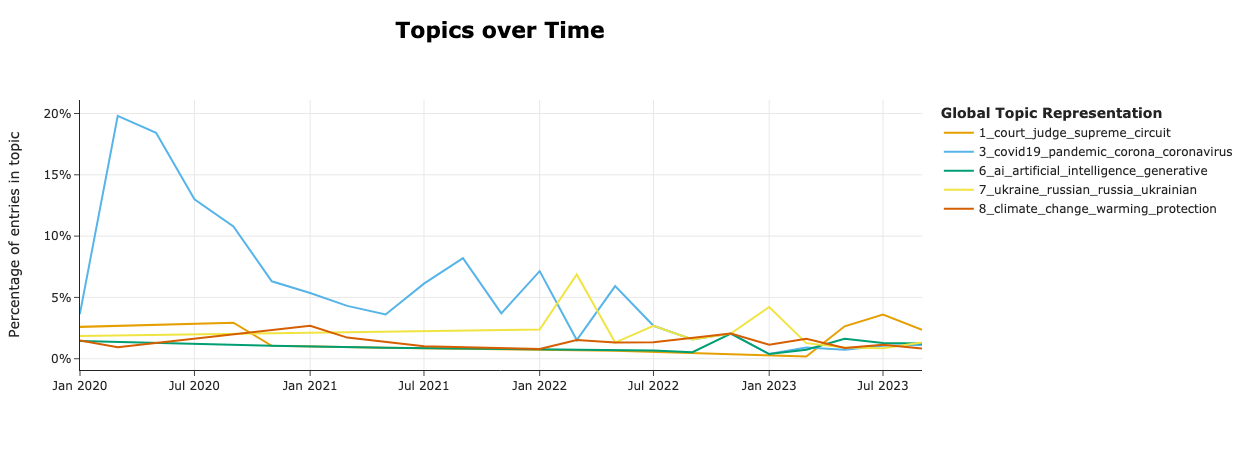}
    \caption{BERTopic visualization of blog entry topics over time. The topics are based on the summaries of the entries. (Interactive plot at \url{https://tobihol.github.io/pollux-rss-blogs/content_analysis/topics_over_time.html})}
    \label{fig:topics_over_time.png}
\end{figure}

\paragraph{Over time visualization.}
We use the publication date of the blog entries to show the trend of topics over time; see \autoref{fig:topics_over_time.png}. The dates are grouped into two-month intervals to make the plot more interpretable. The number of blogs included for dates before the incorporation into Pollux differs from blog to blog, as the number of records per RSS request is individual for each feed, and the frequency of new blogs per feed varies widely. For the visualization, we pick topics that should vary in relevance over time. The blogs cover important political events like the coronavirus pandemic or the war in Ukraine. We see the expected temporal development, where COVID is the dominating topic throughout 2020 and 2021, with events like the Russian invasion of Ukraine spiking at the start of 2022 and the AI-related blogs increasing around the time of the release of ChatGPT in November 2022.

\section*{Acknowledgements}
This work was funded by Deutsche Forschungsgemeinschaft (DFG) under grant number MA 3964/7-2, the POLLUX project.
We thank Marie-Saphira Flug for supporting and reviewing the process of integrating the blog pipeline into Pollux.
We thank \href{https://orcid.org/0000-0002-6656-1658}{\includegraphics[scale=0.06]{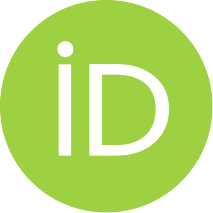}Philipp Mayr} for motivating and supporting the write-up of this technical report.
\printbibliography

\appendix
\section{Representation of Blogs on the Pollux Website}\label{appendix:website}

\begin{figure}[ht]
    \includegraphics[width=\textwidth]{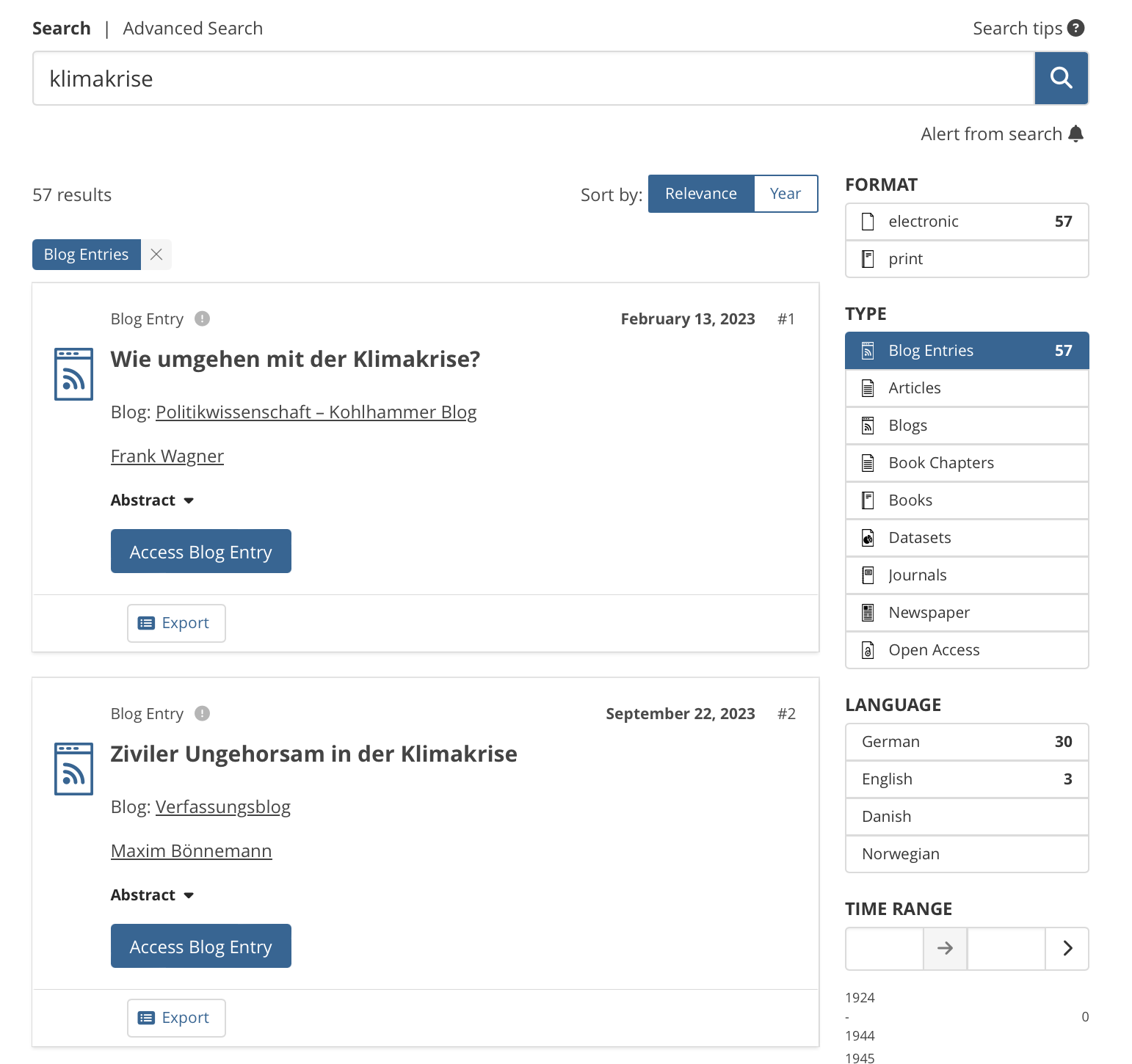}
    \caption{View of the search interface on Pollux when searching for "klimakrise".}
    \label{fig:pollux_web_search_bar}
\end{figure}

\begin{figure}[ht]
    \includegraphics[width=\textwidth]{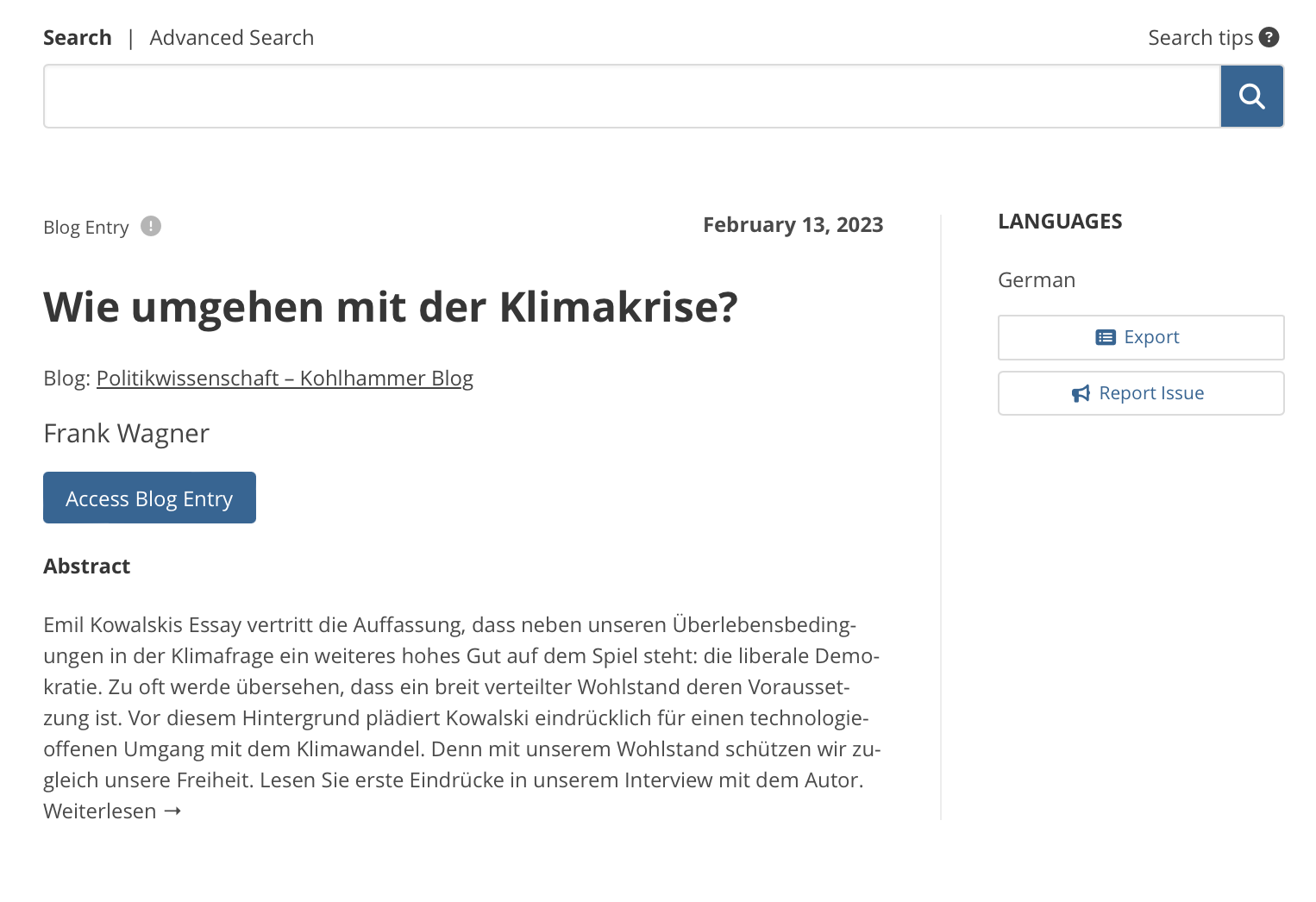}
    \caption{View of a blog entry on Pollux.}
    \label{fig:pollux_web_blog_entry}
\end{figure}

\end{document}